# Gate Switchable Transport and Optical Anisotropy in 90 °Twisted Bilayer Black Phosphorus

Ting Cao[†,‡], Zhenglu Li[†,‡], Diana Y. Qiu[†,‡], Steven G. Louie[*,†,‡]

[†]*Department of Physics, University of California at Berkeley, Berkeley, California 94720, USA.*

[‡]*Materials Sciences Division, Lawrence Berkeley National Laboratory, Berkeley, California 94720, USA.*

[*] sglouie@berkeley.edu

**Abstract**: Anisotropy describes the directional dependence of a material's properties such as transport and optical response. In conventional bulk materials, anisotropy is intrinsically related to the crystal structure, and thus not tunable by the gating techniques used in modern electronics. Here we show that, in bilayer black phosphorus with an interlayer twist angle of 90°, the anisotropy of its electronic structure and optical transitions is tunable by gating. Using first-principles calculations, we predict that a laboratory-accessible gate voltage can induce a hole effective mass that is 30 times larger along one Cartesian axis than along the other axis, and the two axes can be exchanged by flipping the sign of the gate voltage. This gate-controllable band structure also leads to a switchable optical linear dichroism, where the polarization of the lowest-energy optical transitions (absorption or luminescence) is tunable by gating. Thus, anisotropy is a tunable degree of freedom in twisted bilayer black phosphorus.

Atomically thin quasi-two-dimensional (2D) crystals not only unveil a wide range of physical phenomena elusive in their bulk counterparts[1-3], but they may also be constructed into heterogeneous, van der Waals-coupled layers with varied stacking sequences[4] that may exhibit distinct emerging properties. For examples, in graphene-boron nitride[5] and MoS$_2$-MoS$_2$ bilayers[6-9], precise interlayer twist angle control has been achieved, leading to non-local topological currents and unexpected gap evolutions with respect to twist angle. Indeed, material engineering via designed combinations of atomically thin 2D layers is of significant promise in the exploration of novel electronic and optoelectronic applications. Meanwhile, a recently realized 2D crystal, few-layer black phosphorus[10], has attracted a lot of attention due to its anisotropic transport and optical properties[11-15]. Few-layer black phosphorus in its natural stacking order, which has the crystalline orientations of the layers aligned parallel to one another, exhibits similar anisotropy as those of a monolayer[11-13,16]. Here, we explore the effects of changing the relative crystalline orientation of two adjacent layers on the system's anisotropic properties, especially when the relative orientation is at the perpendicular limit (i.e., at 90°).

We performed first-principles calculations on bilayer black phosphorus with an interlayer twist angle of 90° (Figure 1a). We employ a supercell approach[17], taking the in-plane (x-y) dimension of the supercell to be 5×7 times that of a unit cell of monolayer black phosphorus, resulting in nearly commensurate top and bottom layers with a small lattice mismatch of < 1%. The separation between a bilayer and its replica in the neighboring supercell is taken to be 25 Å. The relaxed interlayer distance of the 90° twisted bilayer black phosphorus is 3.3 Å (Figure 1a), which is about 0.2 Å larger than that of the naturally stacked bilayer. The reciprocal space geometry is shown in Figure 1b, in which the rectangular Brillouin Zone (BZ) of the naturally stacked bilayer has been folded onto the square shaped BZ of the supercell structure.

The quasiparticle band structure of the 90° twisted bilayer under zero bias voltage has 4-fold rotational symmetry (nearly isotropic near the Γ point, the BZ center), which is distinct from that of the naturally stacked bilayer. In a naturally stacked bilayer, the quasiparticle band structure only has 2-fold rotational symmetry with a singly degenerate (not counting the spin degree of freedom) conduction band minimum (CBM) and a singly degenerate valence band maximum (VBM) at Γ (Figure 1c). Around the Γ point, the bands are considerably more dispersive along the Γ-X direction than along the Γ-Y direction, corresponding to the armchair direction and zigzag direction in real space, respectively. These singly degenerate CBM and VBM states arise because of interlayer interaction, which splits the otherwise degenerate bands from having two layers. The band splitting between interlayer bonding and anti-bonding orbitals for the CBM and VBM at Γ is 0.4 eV and 0.9 eV, respectively. However, for the 90° twisted bilayer (Figure 1d), although the CBM remains singly degenerate, a two-fold degeneracy appears for the states at the VBM (the energy splitting is < 0.1 meV in our calculation). Moreover, the anisotropic features of the band structure disappear. The energy dispersions along the Γ-X' direction for various bands are identical as those along the Γ-Y' direction. To better illustrate the energy dispersions of bands, we plot the top valence band and the bottom conduction band as energy surfaces in the BZ, shown in the supplementary information section I.

The absence of anisotropy in the 90° twisted bilayer is accompanied by an interesting layer texture of the quasiparticle band states: different bands have distinctly different localization of charge density on the two layers. This is illustrated by a band-index *n* and wavevector $\boldsymbol{k}$ dependent layer localization function $\eta_n(\boldsymbol{k})$, defined as

$$\eta_n(\boldsymbol{k}) = \frac{\Sigma_{i\in t}|\langle\psi_i|\phi_n(\boldsymbol{k})\rangle|^2 - \Sigma_{i\in b}|\langle\psi_i|\phi_n(\boldsymbol{k})\rangle|^2}{\Sigma_{i\in t}|\langle\psi_i|\phi_n(\boldsymbol{k})\rangle|^2 + \Sigma_{i\in b}|\langle\psi_i|\phi_n(\boldsymbol{k})\rangle|^2},$$

where $\phi_n(\boldsymbol{k})$ is the electron wavefunction, and $\psi_i$ denotes an active atomic orbital *i* of a phosphorus atom in the top layer (t) or the bottom layer (b). $\eta_n(\boldsymbol{k})$ quantitatively depicts the spatial distribution of wavefunction of the electronic state, with its value ranging from -1 (i.e., the wavefunction completely localizes in the bottom layer) to 1 (i.e., the wavefunction completely localizes in the top layer). Surprisingly, although $\eta_n(\boldsymbol{k})$ is virtually 0 for all states in naturally stacked bilayer (Figure 1c), it shows large variations for different states in the 90° twisted bilayer (see Figure 1d): the charge density of the states from the lowest unoccupied band near the CBM is equally distributed on both layers; whereas, for the states from the highest occupied band complex, they are separated into two sets of bands – either localized in the top layer (the red bands) or the bottom layer (the blue bands). They are akin to the bands near the VBM of a monolayer; but the two sets of bands (red and blue) are rotated in *k*-space by 90° with respect to each other. We therefore label these two sets of band of different spatial localizations using the nomenclature of VBt and VBb for valence band states in the top and bottom layers, respectively.

The appearance of two identical sets of valence bands, located on the different layers and rotated by 90° in *k*-space, arises from a lack of interlayer hybridization between the states of the two individual layers near the VBM in the 90° twisted structure. Therefore, there are two nearly independent sets of bands that are otherwise identical except for their spatial localization and dispersion. The different behavior between CBM and VBM states in the 90° twisted structure is evident by analyzing the wavefunction symmetry of these states. In Figure 1e and Figure 1f, we plot the wavefunctions of the CBM and the VBM of an isolated monolayer, at a height of 1.6 Å above the phosphorus atoms. This distance corresponds to half of the interlayer separation in the bilayer case, where the wavefunction overlap between layers is most significant. Along the armchair direction (the horizontal direction in Fig. 1e and 1f), the CBM wavefunction (Figure 1f) has the same sign, whereas the VBM wavefunction (Figure 1e) shows alternating positive and negative signs. Consequently, for the CBM states of the two individual layers, their interlayer couplings are always finite regardless of the twist angle. But for the VBM states of the two individual layers, their interlayer coupling become zero as the twist angle approaches 90°. A detailed analysis including the Umklapp processes in a 5×7 supercell is included in the supplementary information section II.

An out-of-plane electric field can shift the energy of the VBt red bands from that of the VBb blue bands by creating a potential difference between the two layers. This gate tunable band structure change leads to a switchable effective mass anisotropy. In our calculation, the effects of gating were simulated by adding a saw-tooth like potential to the supercell including the dipole corrections[18]. In Figure 2a and Figure 2b, we plot the calculated quasiparticle band structure under an out-of-plane electric displacement field of 0.2 V/Å and -0.2 V/Å, as examples of positive and negative bias conditions, respectively. Under positive bias, the VBb complex (blue valence bands) shifts to higher energy compared to the VBt complex (red valence bands). As a result, the hole effective mass along the Γ-X' direction (~ 3.7 $m_0$) is considerably larger than that along the Γ-Y' direction (~ 0.14 $m_0$) at the band edge (i.e., near the new VBM), where $m_0$ is the electron mass in vacuum (Figure 2c). In comparison, as the bias switches to negative, the VBt complex (red valence bands) becomes higher in energy, resulting in a larger hole effective mass along the Γ-Y' direction than along the Γ-X' direction (Figure 2d). For carriers in states near the CBM, however, the electron effective mass remains nearly isotropic and shows less than 20% difference between the two field directions (Figure 2c, d).

In Figure 2e, we plot the energy difference between the VBb and the VBt at Γ as a function of the electric displacement field. The observed linear dependence in the splitting of the two-fold degenerate level is understood as a Stark effect, splitting between two spatially separated nearly degenerate and decoupled states by an applied field. Quite remarkably, the size of the splitting, even at a relatively low field, exceeds the thermal broadening energy at room temperature (26 meV). For example, under an electric displacement field of 0.1 V/Å, the splitting is ~ 50 meV, suggesting measurable anisotropic hole transport at room temperature. In Figure 2f, we compare the bandgap variations of the naturally stacked and 90°twisted bilayer as a function of the electric displacement field. For the naturally stacked bilayer, the bandgap variation shows a quadratic dependence on the field strength; whereas for the 90°twisted bilayer, the bandgap variation shows a close-to-linear dependence mainly due to the Stark splitting of the VBMs. Within the field strength considered in our calculations, the twisted bilayer demonstrates a higher tunability of the bandgap. For example, under a displacement field of 0.2 V/Å, the change of the bandgap in the 90°twisted bilayer, 0.1 eV, is more than 3 times larger than that in the naturally stacked bilayer.

The band structure of the 90°twisted bilayer black phosphorus also brings about two sets of optical transitions, corresponding to transitions from the VBt states to the CB and from the VBb states to the CB (Figure 3a). These two sets of optical transitions are both linearly polarized, but have polarization directions perpendicular to each other. We demonstrate this effect by calculating the polarization direction dependent oscillator strength at the absorption edges, namely $\frac{|\langle \phi_{\text{CB}}|\hat{e}\cdot v|\phi_{\text{VBt}}\rangle|^2}{E_{\text{CB}}-E_{\text{VBt}}}$ and $\frac{|\langle \phi_{\text{CB}}|\hat{e}\cdot v|\phi_{\text{VBb}}\rangle|^2}{E_{\text{CB}}-E_{\text{VBb}}}$, where the unit vector $\hat{e}$ represents the polarization of

the incident light, and $v$ the velocity operator. For optical transitions from VBt-CB, the oscillator strength is consistently 4 orders of magnitude larger for light polarized along the x axis than along the y axis, whereas those from VBb-CB show the opposite behavior. This oscillator strength difference between the two linear polarizations is robust even with an applied electric field. Figure 3b shows an example of the polarization dependent oscillator strength of the two sets of optical transitions under an electric displacement field of 0.2 V/Å, where the two sets of transitions both demonstrate perfect linear dichroism but show orthogonal polarizations.

Owing to the Stark splitting of the VBt and VBb complexes, the excitation energy difference between VBt-CB and VBb-CB transitions can be controlled by applying an electric field. We show the calculated optical absorption spectra with linearly polarized light along the x and y axes under different electric displacement fields in Figure 3c. At zero bias, the absorption spectra of light polarized along x and along y give rise to the same absorption onset at ~ 1.3 eV and are identical. The optical absorbance, within this interband transition framework, is ~ 5% above the absorption onset, and shows a step-function feature with excitation energy, originating from the joint density of states of a direct-bandgap 2D semiconductor. Applying an electric field breaks the degeneracy between the two sets of transitions, leading to higher energy absorption onset for light polarized along one axis than along the other axis (Figure 3c). Therefore, by tuning the direction of the applied electric field, the polarization of the lowest energy optical transition can be switched between two orthogonal directions. Under the electric displacement field of 0.2 V/Å, the onset energy difference between the two spectra is 0.1 eV, which is equal to the Stark splitting of the two band complexes under the same field strength (Figure 2e). Here we have not included excitonic effects in the discussion; but such effects would not change the main conclusions since the lowest energy excitons are linear combination of interband transitions of either VBt-CB or VBb-CB types with weak mixing between them because of the spatial separation of VBt and VBb hole states from each other.

The 90° twisted bilayer black phosphorus may also be used as a platform for polarization dependent photovoltaic effects. Considering light polarized along the x axis which excites the VBt-CB transitions, the generated photoelectrons at the CB are distributed equally in both layers, whereas the holes at the VBt are mostly localized in the top layer. This imbalanced charge carrier distribution would lead to a transient voltage drop from the top layer to the bottom layer. As this effect originates from the polarization-dependent linear dichroism, the photovoltage between the two layers can switch from positive to negative, when the polarization direction rotates from the x axis to the y axis. We estimate that the maximum size of the transient voltage from a photocarrier density of $10^{13}$/cm$^2$ is ~0.03 V, if light is polarized along one axis (see supplementary information section III for details).

Experimentally, samples of 90° twisted bilayer black phosphorus may be realized by stacking two monolayers with perpendicular crystalline orientations. The electric displacement field can be controlled by using a dual-gate setup, for which earlier experiments have demonstrated a tunable field strength up to ~ 0.3 V/Å in bilayer graphene[19]. Meanwhile, as ambipolar and high mobility transport have been achieved in black phosphorus field effect transistors[10,20], we expect that the gate-switchable, anisotropic hole effective mass may be confirmed by transport measurements with a dual-gate setup. The switchable optical linear dichroism may also be detected in polarization dependent absorption measurements.

As a perfect alignment of the two layers at exactly 90° from each other might be difficult to achieve experimentally, we also performed calculations on bilayers with an 82° twist angle to investigate the robustness of the field switchable anisotropy with a certain misalignment. For an 82° twisted bilayer, the calculated band structure shows a small splitting of ~8 meV of the otherwise two-fold degenerate VBM at 90°. This result indicates a finite but very weak interlayer coupling of the valence band wavefunctions, when the interlayer twist angle deviates from 90°. However, the small splitting of ~8 meV at 82° is still much smaller than the field induced Stark splitting of the VBMs separating the VBt from the VBb, which can be as large as ~100 meV under a displacement field of 0.2 V/Å. Therefore, the predicted switchable anisotropy is expected to be robust against an interlayer twist angle misalignment that is ~10°.

As the coupling between the VBM wavefunctions of two native bilayers at 90°, in principle, has the same symmetry property as that between two monolayers, the field switchable anisotropy predicted here also applies to the stacking of two native bilayers.

In summary, we demonstrate gate-switchable effective mass anisotropy and gate-switchable optical linear dichroism in 90° twisted bilayer black phosphorus from first-principles calculations. Our study suggests that anisotropy in stacked two-dimensional crystals may serve as a tunable degree of freedom for their future electronic and optoelectronic applications, and largely increases the functionality of black phosphorus and a range of other 2D materials where anisotropy exists, such as Re-based chalcogenides[21], as well as Sn- and Ge- based monochalcogenides[22].

**Figures and Captions:**

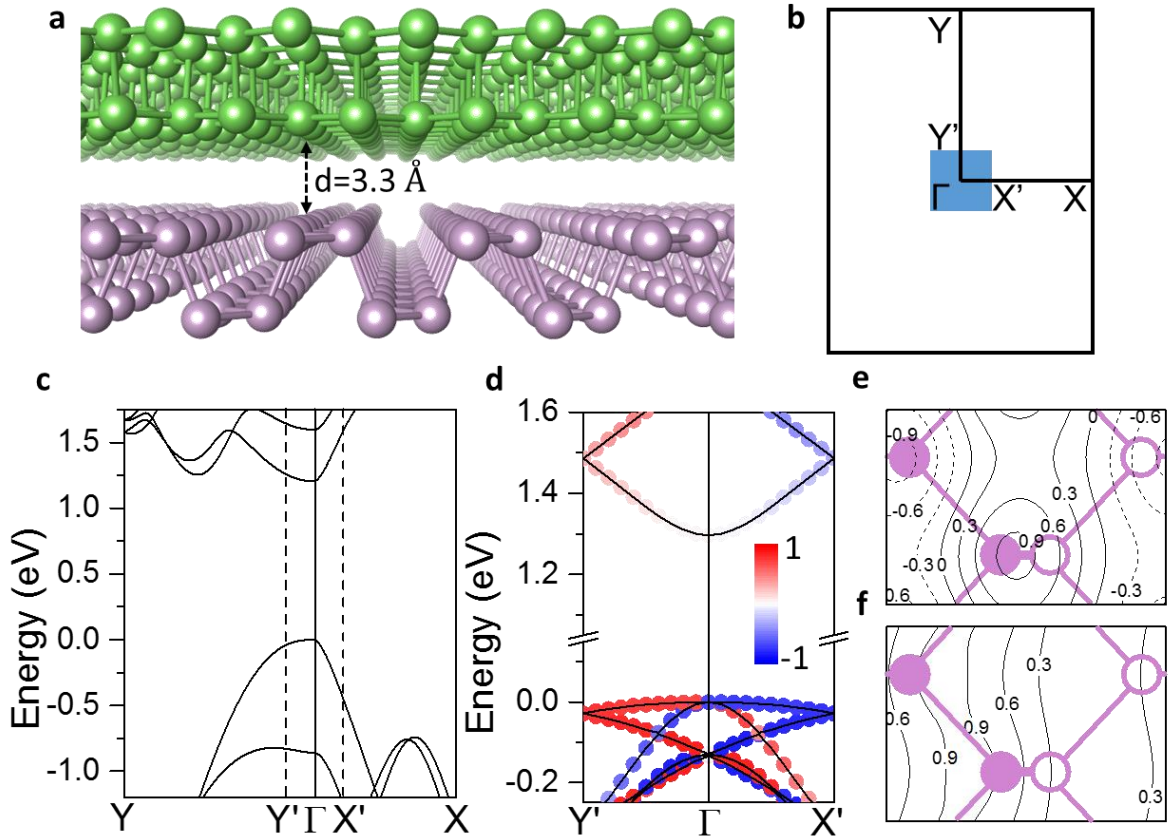

**Figure 1.** Structure and electronic structure of 90° twisted bilayer black phosphorus. (a) Structure of a 90° twisted bilayer black phosphorus. The phosphorus atoms in the top and bottom layers are colored by cyan and purple, respectively. The interlayer distance is 3.3 Å from first-principles DFT-PBE calculation with van der Waals correction. (b) Schematic of the Brillouin zone (BZ) of bilayer black phosphorus in its natural stacking order (rectangle) and the BZ of a 5x7 supercell of 90° twisted bilayer black phosphorus (shaded square). (c) GW quasiparticle band structure of naturally stacked bilayer black phosphorus. (d) GW quasiparticle band structure of 90° twisted bilayer black phosphorus, with the color representing the layer localization function defined in the main text. (e,f) Contour plots of the wavefunctions of (e) the valence band maximum and (f) the conduction band minimum of a monolayer black phosphorus at a height of 1.6 Å above the phosphorus atoms, respectively. The wavefunction is normalized to its maximum amplitude in each plot. The solid and dashed lines represent positive and negative values of the wavefunctions, respectively. The solid and hollow purple circles denote the phosphorus atoms at the top and bottom of a monolayer, respectively.

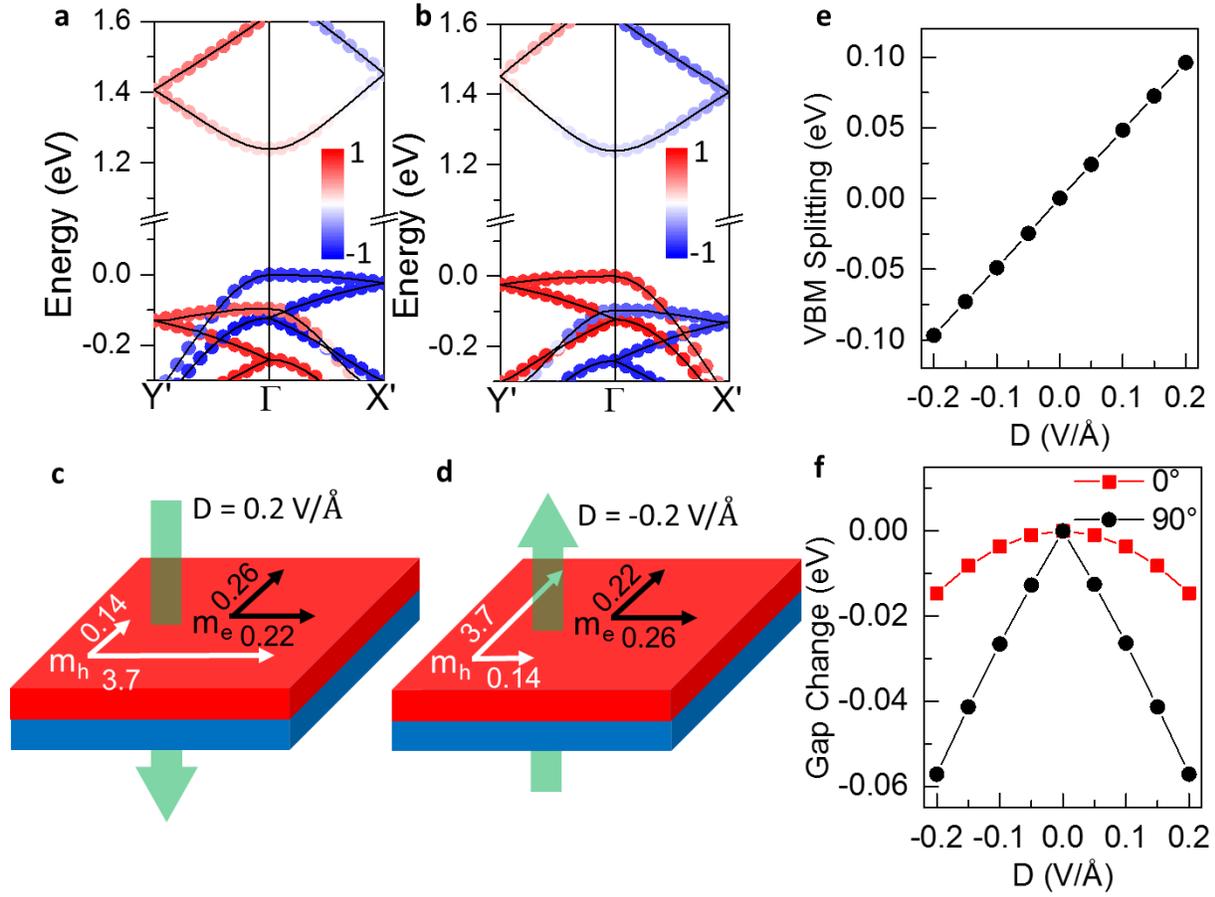

**Figure 2.** Electronic structure of the 90° twisted bilayer black phosphorus under out-of-plane electric displacement field. (a,b) GW quasiparticle band structure of the 90° twisted bilayer black phosphorus in a perpendicular electric displacement field of 0.2 V/Å and -0.2 V/Å, respectively. The color represents the layer localization function defined in the main text. (c,d) Schematics of effective mass under electric displacement field of (c) 0.2 V/Å and (d) -0.2 V/Å. The hole and electron effective masses are shown in white and black, respectively. The effective masses are in unit of the bare electron mass. The green arrow denotes the direction of the displacement field. (e) Splitting of the two-fold degenerate valence band maximum (VBM) as a function of the electric displacement field. The lines are visual guides. (f) The change in bandgap as a function of the electric displacement field. Red squares and black circles denote the bandgap change in 0° naturally stacked bilayer and 90° twisted bilayer, respectively. The lines are visual guides.

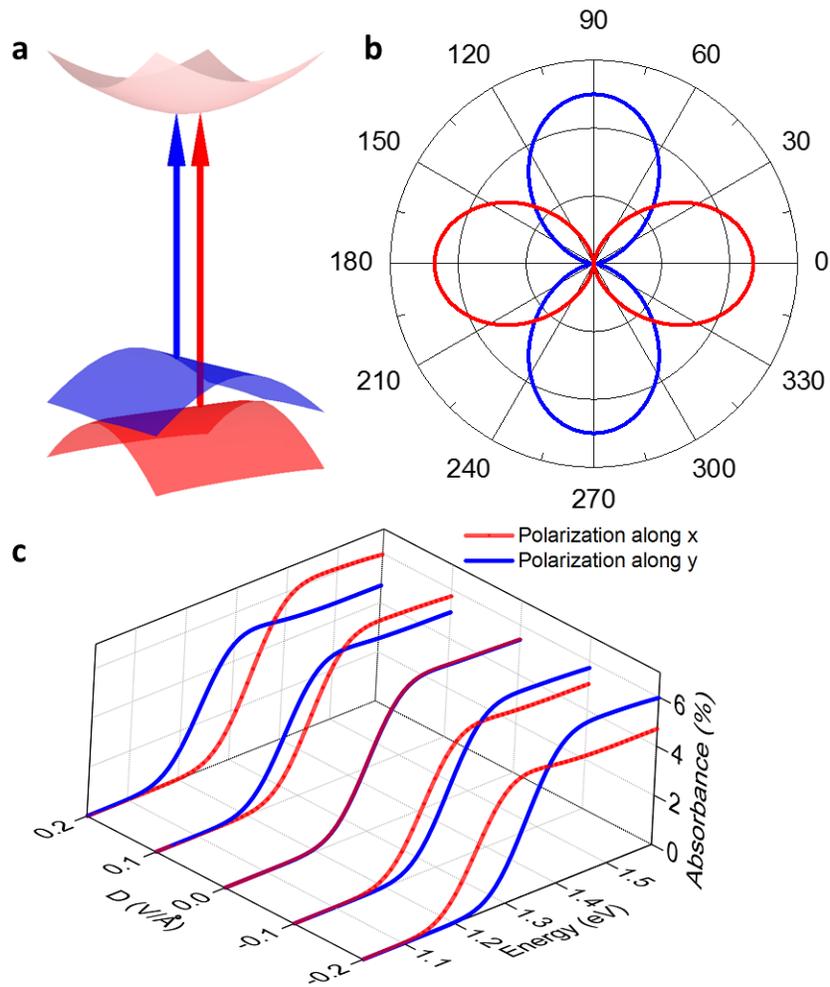

**Figure 3.** Optical transitions of the 90° twisted black phosphorus bilayer under different electric displacement field from first principles. (a) Schematic of optical transitions from VBb states to CB states (blue arrow) and VBt states to CB states (red arrow) under the electric displacement field of 0.2 V/Å. (b) Oscillator strength of optical transitions from VBb-CB (blue) and VBt-CB (red) at Γ as a function of the direction of the light polarization. The oscillator strength is plotted in arbitrary unit as a function of the polarization angle, with 0° along the x axis, and 90° along the y axis. (c) Optical absorption spectra of linearly polarized light along x axis (red) and y axis (blue) under electric displacement field from -0.2 V/Å to 0.2 V/Å.

**Method:**

First-principles calculations were performed using density functional theory in the generalized gradient approximation (GGA) implemented in the Quantum Espresso package[23]. The quasiparticle band structures were calculated with the *ab initio* GW method[24] as implemented in the BerkeleyGW package[25]. A supercell arrangement was used, with the vacuum thickness set to 18 Å and a truncated Coulomb interaction along the out-of-plane direction to avoid interactions between the black phosphorus bilayer and its periodic images. We employed norm-conserving pseudopotentials, with a plane-wave energy cutoff of 40 Ry. In the relaxation of the structures, we included dispersion corrections within the D2 formalism to account for the van der Waals interactions[26]. The structure was fully relaxed until the force on each atom was smaller than 0.01 V/Å. The effects of gating were simulated by adding a saw-tooth like potential to the supercell including the dipole corrections. The size of the out-of-plane electric displacement field was self-consistently fixed in each calculation. In calculating the quasiparticle band structures of the supercell, we employed a rigid-band approximation in the construction of the Green's function. The conduction band Kohn-Sham eigenvalues are shifted collectively by the amount of self-energy corrections to the bandgap at the $\Gamma$ point. The dielectric matrix was constructed with a cutoff energy of 8 Ry and 10,000 empty bands. The dielectric matrix was calculated on a 4×4×1 coarse k-grid with subsampled fine grids for q-point approaches 0. The quasiparticle bandgap is converged within 0.1 eV. The optical transitions were calculated using the linear response theory under the dipole approximation[27] implemented in the BerkeleyGW package, with a Gaussian broadening of 0.05 eV. The transition energies are derived from the quasiparticle band structure. Excitonic effects were not included.